# Kinetics of binding and geometry of cells on molecular biochips

V.R. Chechetkin[*]

*Theoretical Department of Division for Perspective Investigations, Troitsk Institute of Innovation and Thermonuclear Investigations (TRINITI), Troitsk,
142190 Moscow Region, Russia*

**Abstract**

We examine how the shape of cells and the geometry of experiment affect the reaction-diffusion kinetics at the binding between target and probe molecules on molecular biochips. In particular, we compare the binding kinetics for the probes immobilized on surface of the semispherical and flat circular cells, the limit of thin slab of analyte solution over probe cell as well as hemispherical gel pads and cells printed in gel slab over a substrate. It is shown that hemispherical geometry provides significantly faster binding kinetics and ensures more spatially homogeneous distribution of local (from a pixel) signals over a cell in the transient regime. The advantage of using thin slabs with small volume of analyte solution may be hampered by the much longer binding kinetics needing the auxiliary mixing devices. Our analysis proves that the shape of cells and the geometry of experiment should be included to the list of essential factors at biochip designing.

## 1. Introduction

The operation principle of molecular biochips is based on the highly specific molecular recognition between the target molecules in analyte solution and the probe molecules immobilized either in the cells on a surface or within gel pads on a substrate. In the common architecture the cells form $L \times M$ matrix comprising $\sim 10^3$-$10^5$ individual patterns, where the each cell in the matrix is designated for recognition of particular target molecules. The characteristic radii of cells in the modern biochips are about ~100 μm, while the characteristic lateral sizes of biochips amount to ~1 cm and the height of hybridization chamber ~100 μm. These miniature tools serve for the high-throughput analysis in a variety of genetic, biomedical, and ecological problems [1-5]. Among other numerous factors, the operation efficiency and sensitivity of such devices depends on the shape of cells and geometry of hybridization chamber [6]. In this paper we will discuss how the shape of cells and the geometry of experiment affect the reaction-diffusion binding kinetics and the distribution of local signals on the cells in the transient regime.

## 2. General equations

The evolution of the volume concentration of the target molecules in solution, $h(\mathbf{r}, t)$, is governed by the diffusion equation

$$\frac{\partial h}{\partial t} = D_{sol} \nabla^2 h \tag{1}$$

---

[*] *E-mail addresses:* chechet@biochip.ru and vladimir_chechet@mail.ru.

where $D_{sol}$ is diffusion coefficient for target molecules in solution. At a boundary of a cell with the immobilized probes the balance related to the formation and dissociation of target-probe complexes is determined by diffusive flow,

$$\frac{\partial \tilde{d}}{\partial t} = -D_{sol} \mathbf{n}_b \cdot \nabla h \Big|_{\mathbf{r} \in cell} = k_{ass}(\tilde{m} - \tilde{d}) h \Big|_{\mathbf{r} \in cell} - k_{diss} \tilde{d} \qquad (2)$$

while at a surface beyond a cell the diffusive flow should be equal to zero

$$D_{sol} \mathbf{n}_b \cdot \nabla h \Big|_{\mathbf{r} \in surface;\ \mathbf{r} \notin cell} = 0 \qquad (3)$$

Here $\mathbf{n}_b$ is the unit vector normal to a boundary of a substrate surface and directed from solution to substrate, $\tilde{m}$ is the density of the immobilized probes per unit area, $\tilde{d}$ is the surface density of the formed target-probe complexes, while $k_{ass}$ and $k_{diss}$ are the association and dissociation rates respectively. Eqs. (1) and (2) ought to be solved with the initial conditions

$$\tilde{d}(\mathbf{r},t)\Big|_{\mathbf{r} \in cell,\ t=0} = 0 \qquad (4)$$

Additionally, $h(\mathbf{r},t)$ should tend to the volume concentration of target molecules $h_{sol}$ far away from a cell.

The current local observable signal from the cell, $J(\mathbf{r}, t)$, is assumed to be linearly related to the formed target-probe complexes,

$$J(\mathbf{r},t) = A\,\tilde{d}(\mathbf{r},t) \qquad (5)$$

where $A$ is apparatus constant.

The characteristic time related to the smoothing of concentration variations in the outer region from a cell is about $\sim R^2/D_{sol}$, while the characteristic time needed for the saturation of binding on the probe cell $\tau_B$ (and respectively for the saturation of observable signal from the cell) is usually much longer. Thus, all the consideration below implies the fulfillment of conditions,

$$t \gg R^2/D_{sol},\ \tau_B \gg R^2/D_{sol} \qquad (6)$$

The inequalities (6) should be taken into account at the extrapolation of solutions presented below to the initial moment $t = 0$.

## 3. Hemispherical geometry of surface cell

At $t \gg R^2/D_{sol}$ the asymptotic distribution of concentration of the target DNA in solution beyond hemispherical cell region can be approximated with a reasonable accuracy by quasistationary profile,

$$h(r,t) = \big(h(t) - h_{sol}\big)\frac{R}{r} + h_{sol} \qquad (7)$$

where $h_{sol}$ is the homogeneous concentration of target molecules in solution far away a cell with immobilized probes, $h(t)$ corresponds to the concentration of target molecules at the cell surface, and $r$ is spherical radial coordinate.

The substitution of concentration (7) into Eq. (2) yields the relationship

$$-D_{sol} \mathbf{n}_b \cdot \nabla h \big|_{\mathbf{r} \in cell} \cong \frac{D_{sol}}{R} \left( h_{sol} - h(t) \right) \tag{8}$$

After that, the explicit integration provides the evolution of fluorescence signal

$$(1-\eta_{eq})\left(1 + \frac{1}{\tau_{B,diff}^{(sphere)} k_{diss}}\right) \ln\left(1 - \frac{J(t)}{J_{eq}}\right) - \eta_{eq} \frac{J(t)}{J_{eq}} = -\frac{t}{\tau_{B,diff}^{(sphere)}} \tag{9}$$

Here $J(t)$ is current fluorescence signal evolving to its saturation value $J_{eq}$ at thermodynamic equilibrium,

$$J_{eq} = A \tilde{m} \eta_{eq} \tag{10}$$

and

$$\eta_{eq} = \frac{K_a h_{sol}}{1 + K_a h_{sol}} \tag{11}$$

$$\tau_{B,diff}^{(sphere)} = \frac{R \tilde{m} K_a}{D_{sol}(1 + K_a h_{sol})} \tag{12}$$

where $K_a = k_{ass}/k_{diss}$ is thermodynamic association constant. The factor $\eta_{eq}$ characterizes the equilibrium fraction of probes participating in the formation of target-probe complexes. The observable signal is spatially homogeneous throughout cell area. Eq. (9) coincides formally with the evolution equation for the binding kinetics on the flat cells derived in Refs. [7, 8] using two-compartment approximation.

The physical meaning of Eq. (12) becomes clearer if it would be rewritten in the form

$$2\pi R h_{sol} D_{sol} \tau_{B,diff}^{(sphere)} = 2\pi R^2 \tilde{m} \frac{K_a h_{sol}}{(1 + K_a h_{sol})} \tag{12'}$$

The left hand side of Eq. (12′) corresponds to the diffusion flux accumulating the target molecules on the cell surface, while the right hand side denotes the total number of complexes which may be formed between target and probe molecules in thermodynamic equilibrium. The balance between two factors is needed for the saturation of binding on hemispherical cell.

The parameter $\tau_{B,diff} k_{diss}$ determines the transition from diffusion-limited (if $\tau_{B,diff} k_{diss} \gg 1$) to reaction-limited (in the opposite case $\tau_{B,diff} k_{diss} \ll 1$) kinetics. In the latter case the characteristic time for binding saturation is given by

$$\tau_{B,react} = \frac{1}{k_{diss} + k_{ass} h_{sol}} \tag{13}$$

and does not depend on geometry.

## 4. Flat circular geometry of surface cell

The respective kinetics for the flat circular geometry may be obtained by extending Weber's solution for the stationary accretion of particles on a disc [9] to binding problem (the relevance of this solution to binding kinetics as well as to the inhomogeneity of signal distribution over flat circular cells has been remarked in Refs. [6, 10]). Formally, this solution also describes the distribution of potential around charged metallic disc [11]. The relevant extension of Weber's solution to the problem of binding kinetics has the form

$$h(r, z, t) = \frac{2}{\pi}(h(t) - h_{sol}) \tan^{-1}\left[\frac{2R^2}{r^2 + z^2 - R^2 + [(r^2 + z^2 - R^2)^2 + 4R^2 z^2]^{1/2}}\right]^{1/2} + h_{sol} \quad (14)$$

Here $r$ and $z$ denote cylindrical coordinates with the axis on the center of the cell and $\tan^{-1}$ means arc tangent. Distribution (14) provides the spatially inhomogeneous diffusion flux on the cell surface,

$$D_{sol} \left.\frac{\partial h}{\partial z}\right|_{z=0, r \leq R} \cong \frac{2}{\pi} \frac{D_{sol}}{\sqrt{R^2 - r^2}} (h_{sol} - h(t)) \quad (15)$$

The evolution of a local (from a pixel) signal is given by

$$(1 - \eta_{eq})\left(1 + \frac{1}{\tau_{B, diff}^{(disc)}(r) k_{diss}}\right) \ln\left(1 - \frac{J(r, t)}{J_{eq}}\right) - \eta_{eq} \frac{J(r, t)}{J_{eq}} = -\frac{t}{\tau_{B, diff}^{(disc)}(r)} \quad (16)$$

where

$$\tau_{B, diff}^{(disc)}(r) = \frac{\pi}{2} \frac{\sqrt{R^2 - r^2}}{D_{sol}} \frac{\tilde{m} K_a}{(1 + K_a h_{sol})} \quad (17)$$

At the same radius of hemispherical and flat circular cells $R$, beginning from $r < 0.77 R$, the binding kinetics on disc becomes slower than that on hemispherical cell. The kinetics of signal saturation at the center of flat cell turns out approximately 1.6-fold slower than the rate of binding at hemisphere. The averaging of time (17) over the cell surface provides

$$<\tau_{B, diff}^{(disc)}> = \frac{\pi}{3} \frac{R}{D_{sol}} \frac{\tilde{m} K_a}{(1 + K_a h_{sol})} \quad (18)$$

which is about 10% higher than (12) (at the twice larger surface area for hemisphere at the same radius $R$). Unlike hemispherical geometry, the signal for flat cells is strongly spatially inhomogeneous in the transient regime of binding.

The radial dependence of time (17) leads to the reaction-limited kinetics of binding at the periphery of flat circular cell in the region $r \approx R$. In the practice this region may be quite narrow. Specifically, if $\tau_{B, diff}^{(disc)}(r = 0) k_{diss} = 5$ (or even larger in many cases), then the transition to the reaction-limited kinetics takes place from $r > 0.87R$.

Fig. 1 illustrates the comparison of binding kinetics on hemispherical and flat circular cells. These kinetic curves show that the difference in times needed for saturation of signals at the periphery and at the center of flat circular cell may be quite significant.

The distributions (7) and (14) obtained for the bulk geometry can be straightforwardly generalized to the practically important situation with finite relatively thick slab of height $H > R$ of analyte solution over a cell by the standard technique of iterative mirror reflections with respect to both slab surfaces. The effects of finite slab height are determined by the parameter $R/H$ (typically, ~0.1–0.5). The leading corrections on the parameter $R/H$ to the binding time (17) are given by

$$\tau_{B,diff}^{(disc)}(r, H) \cong \tau_{B,diff}^{(disc)}(r)\left(1 + \frac{2R}{\pi H}\ln\left((\tilde{m}K_a R)^{1/2}/H\right)\right) \tag{19}$$

Eq. (19) implies the fulfillment of inequality $(\tilde{m}K_a R)^{1/2} > H$, otherwise the finite height corrections may be neglected. These corrections illustrate the general influence of finite slab height. The thinner the slab of analyte solution over a cell, the slower the binding kinetics needed for signal saturation. In the case of hemispherical geometry the effects of finite slab height also cause the violation of the spatial homogeneity of signal distribution over a cell (though in the second order on $R/H$). The limit of thin slab of analyte solution over a cell will be considered in the next section.

## 5. Thin slab of analyte solution over flat circular cell

In some cases it is desirable to work with volume of analyte solution as small as possible by thinning slab of solution over a cell. The thin slabs are determined by the condition $\delta \ll R$, where $\delta$ is thickness of analyte solution slab over a cell with radius $R$ (typically, $\delta \sim 20\text{-}30$ μm, while $R \sim 100\text{-}200$ μm). It is assumed that $\tilde{m}K_a/\delta \gg 1$ and $K_a h_{sol} \sim 1$. In these conditions the target molecules will be supplied to a cell by the radial diffusion flow from the outer region beyond a cell. In the outer region the transverse variations of concentration along height of slab are much less than the counterpart radial variations. For simplicity, the distribution of concentration will be considered as homogeneous along height of slab (or instead, the concentrations may be integrated over a height of slab). Consider, first, the radial distribution of diffusion flux in the outer region with the concentration $h(R, t)$ quasi-statically held fixed at the boundary of cell, $r = R$ (physically, the fixation of $h(R, t)$ means that the characteristic time $\tau_{B,diff}$ for the changes in $h(R, t)$ exceeds significantly the diffusion time $\sim R^2/D_{sol}$). At $t \gg R^2/D_{sol}$ and $r > R$ the outer radial distribution of concentration in this geometry may be approximated as

$$h(r,t) = (h(R,t) - h_{sol})\frac{\ln(R_{max}/r)}{\ln(R_{max}/R)} + h_{sol} \tag{20}$$

with $R_{max} \cong \min\{(D_{sol} t)^{1/2}, L_{chip}\}$. Here $L_{chip}$ corresponds to the characteristic lateral size of a chip. At $r > R_{max}$ it should be taken $h(r,t) \cong h_{sol}$. Choosing for the assessment of $R_{max}$ the typical parameters $D_{sol} \sim 10^{-6}$ cm$^2$/s, $t \sim 10^4$ s and $L_{chip} \sim 1$ cm, one obtains the value of $R_{max} \cong (D_{sol} t)^{1/2}$. The estimate for the incoming radial flux turns out about

$$D_{sol}\frac{\partial h}{\partial r}\bigg|_{r=R} \cong \frac{2 D_{sol}}{R\ln(D_{sol} t / R^2)}(h_{sol} - h(R,t)) \tag{21}$$

The more rigorous approach is based on Laplace transform for the concentration distribution

$$h(r, p) \cong \frac{h_{sol}}{p} + \frac{(h(R, t) - h_{sol})}{p} \frac{K_0(r(p/D_{sol})^{1/2})}{K_0(R(p/D_{sol})^{1/2})}, \quad R \leq r < \infty \tag{22}$$

$$h(r, t) = \frac{1}{2\pi i} \int_{a-i\infty}^{a+i\infty} dp\, e^{pt}\, h(r, p), \quad a > 0 \tag{23}$$

Here $K_0(x)$ is Macdonald function (or modified Bessel function from imaginary argument) with the zeroth index. Then, the consideration similar to Ref. [12] gives the relationship,

$$D_{sol} \left.\frac{\partial h}{\partial r}\right|_{r=R} \cong \frac{2 D_{sol}}{R \ln(\pi D_{sol} t / R^2 \gamma^2)} (h_{sol} - h(R, t)) \approx \frac{2 D_{sol}}{R \ln(1.26 D_{sol} t / R^2)} (h_{sol} - h(R, t)) \tag{24}$$

practically coincident with the above approximation (21). Here $\gamma = e^C \approx 1.7810...$ ($C$ is Euler constant). Note that the logarithmic suppression of the radial diffusion flux (24) with time $t$ is the generic feature of diffusion-limited association processes in two-dimensional systems [13–15].

Due to the conservation law, the radial diffusion flux at the vicinity of a cell and the transverse diffusion on a cell should be related by the balance equation,

$$D_{sol} 2\pi R \delta \left.\frac{\partial h}{\partial r}\right|_{r=R} \cong D_{sol} 2\pi \int_0^R dr\, r \left.\frac{\partial h}{\partial z}\right|_{z=0} = 2\pi \frac{\partial}{\partial t} \int_0^R dr\, r\, \tilde{d}(r, t) \tag{25}$$

In the region over the cell $0 \leq r \leq R$ the distribution of concentration may be approximated as

$$h(r,t) = h(R, t) \frac{\ln(r/r(t))}{\ln(R/r(t))}, \quad r(t) \leq r \leq R; \tag{26}$$
$$h(r,t) = 0, \quad 0 \leq r \leq r(t)$$

where $r(t)$ corresponds to the characteristic front of expanding binding. The continuity of radial flux at $r = R$ provides the relationship between $r(t)$ and $h(R, t)$,

$$D_{sol} \left.\frac{\partial h}{\partial r}\right|_{r=R} \cong \frac{2 D_{sol}}{R \ln(1.26 D_{sol} t / R^2)} (h_{sol} - h(R, t)) = \frac{D_{sol} h(R, t)}{R \ln(R/r(t))} \tag{27}$$

For the diffusion-limited binding kinetics the local quasi-equilibrium for the formed complexes may be assumed

$$\tilde{d}(r, t) \approx \tilde{m} \frac{K_a h(r, t)}{1 + K_a h(r, t)} \tag{28}$$

Combining Eqs. (26)–(28) with Eq. (25) yields the evolution equation for $r(t)$,

$$\frac{2\delta D_{sol} h_{sol}}{2\ln(R/r(t)) + \ln(1.26 D_{sol} t / R^2)} = \frac{\partial}{\partial t} \int_{r(t)}^R dr\, r\, \tilde{d}(r, t) \tag{29}$$

At the initial stage of binding, when $\tau_B^{(slab)} \gg t \gg R^2/D_{sol}$ (here $\tau_B^{(slab)}$ is defined by Eq. (31) below), the evolution of *r(t)* is approximately diffusion-like

$$(R-r(t))^2 \approx \frac{2\delta D_{sol} t}{\tilde{m} K_a} \tag{30}$$

As is seen from Eq. (30), the effective diffusion coefficient for the evolution of *r(t)*, $D_{eff}^{(slab)} \cong \delta D_{sol}/\tilde{m} K_a$, turns out much lower than $D_{sol}$. The solution of Eq. (29) determines the characteristic time needed for binding saturation in this geometry,

$$\tau_{B, diff}^{(slab)} \cong \frac{R}{\delta} \tau_{B, diff} \ln\left(D_{sol} \tau_{B, diff}/R\delta\right) \tag{31}$$

Here the binding time $\tau_{B, diff}$ is defined by Eq. (12). The binding kinetics in such geometry is actually always diffusion-limited. As $R \gg \delta$, $\tau_{B,diff} \gg R^2/D_{sol}$, the characteristic time (31) exceeds significantly $\tau_{B, diff}$. Therefore, the advantage of using thin slabs with small volume of analyte solution may be hampered by the much slower binding kinetics.

## 6. Gel pads

In the molecular biochips with the gel pads the probes are immobilized either in the hemispherical gel pads or printed in gel slab over a substrate. The characteristic time for the binding kinetics on the gel pads with radius *R* is always diffusion-limited and is about [16]

$$\tau_{B, diff}^{(gel)} \cong \frac{R^2 m K_a}{D_{gel}(1 + K_a h_{sol})} \tag{32}$$

where *m* now means a concentration of probes in a gel pad and $D_{gel}$ corresponds to the diffusion coefficient for diffusion of target molecules in gel without probes. Let the total number of probes immobilized in surface and gel cells of identical radius *R* be comparable

$$\pi R^2 \tilde{m} \approx \frac{2}{3}\pi R^3 m \tag{33}$$

Juxtaposing characteristic times (12) and (32) at the condition (33) reveals that the binding kinetics in the gel pads is slower with respect to that on the surface cells by the factor $\sim D_{gel}/D_{sol}$. This slowing down of the binding kinetics is, however, compensated by the advantage of much longer separation distances for the probes immobilized in gel pad with respect to the counterpart distances in surface cell because the condition (33) proves that,

$$\tilde{m}^{-1/2} \ll m^{-1/3} \tag{34}$$

The immobilization with surface density $\tilde{m} = 10^{13}$ molecules/cm$^2$ would correspond to the distance between probes about 32 Å = 3.2 nm, which is comparable or commonly less than the characteristic sizes (~50-100 Å) of relatively large biological macromolecules used for the probes. Therefore, the immobilization with such surface density would hamper the molecular interactions between target and probe molecules and diminishes the accessibility of probes. Besides that, the substrate surface introduces the additional steric restrictions to the molecular interactions between target and probe molecules [17]. Unlike the surface cells, the immobilization of the same number of probes within gel pad with radius $R$ = 100 μm would provide the interprobe distance about 870 Å = 87 nm, which is much larger than the sizes of macromolecules and does not introduce any restrictions for the accessibility of probes. This store in separation distance between probes allows also to enhance the number of immobilized probes within gel pads and to increase the related observable signals from them (see also [3, 8]).

The similar comparison between times (31) and (32) demonstrates that in the case of thin slab of analyte solution over flat circular cell the factor $(R/\delta) \ln(D_{sol} \tau_{B,diff} / R\delta)$ may significantly exceed the ratio $D_{sol} / D_{gel}$ characterizing the slowing down of the diffusion in gel with respect to that in analyte solution. Really, systems with thin slab geometry cannot be used efficiently without auxiliary mixing devices.

Extending above results to the comparison between the hemispherical gel pads and the cells printed within flat gel slab proves that the diffusion fluxes for hemispherical pads should provide faster binding kinetics and ensure more spatially homogeneous signals on hemispherical pads in the transient regime.

## 7. Conclusion

As in the practice the time of analysis tends to be shortened, the most of signal measurements from molecular biochips are performed in the transient regime far from thermodynamic equilibrium. In such regime the kinetic effects play important role in the discrimination between specific and non-specific binding [18], which is crucial for the applications of biochips. The kinetic effects also strongly influence the choice of optimal parameters and the sensitivity of biochips [6–8]. Our analysis shows that the shape of cells and the geometry of experiment should be included to the list of essential factors at biochip designing. For simplicity we restricted ourselves to single-analyte solutions. In the case of multi-analyte solutions the effects of competitive hybridization must be taken into account [19–22]. If necessary, they may easily be incorporated in the above scheme.

The data processing from molecular biochips needs the analysis of huge massive of noisy data [23]. Therefore, such factors as spatial homogeneity of signal distribution over a cell cannot be neglected as well. In this sense the hemispherical geometry turns out the most optimal from the point of view of both binding kinetics and homogeneity of signal distribution. At the advance in modern technology producing cells of hemispherical shape does not create any principal difficulties.

The application of active mixing for the acceleration of external transport mitigates partially the effects of cell shape. Yet even in this case the viscous forces lead to the zero transport velocity at substrate surface and retain the essential contribution of diffusive fluxes to binding kinetics. For this reason, the role of geometric factors is worth having in mind in the further optimization of biochip efficiency.

**Acknowledgements**

The author is grateful to A. Turygin and D. Prokopenko for helpful discussions.

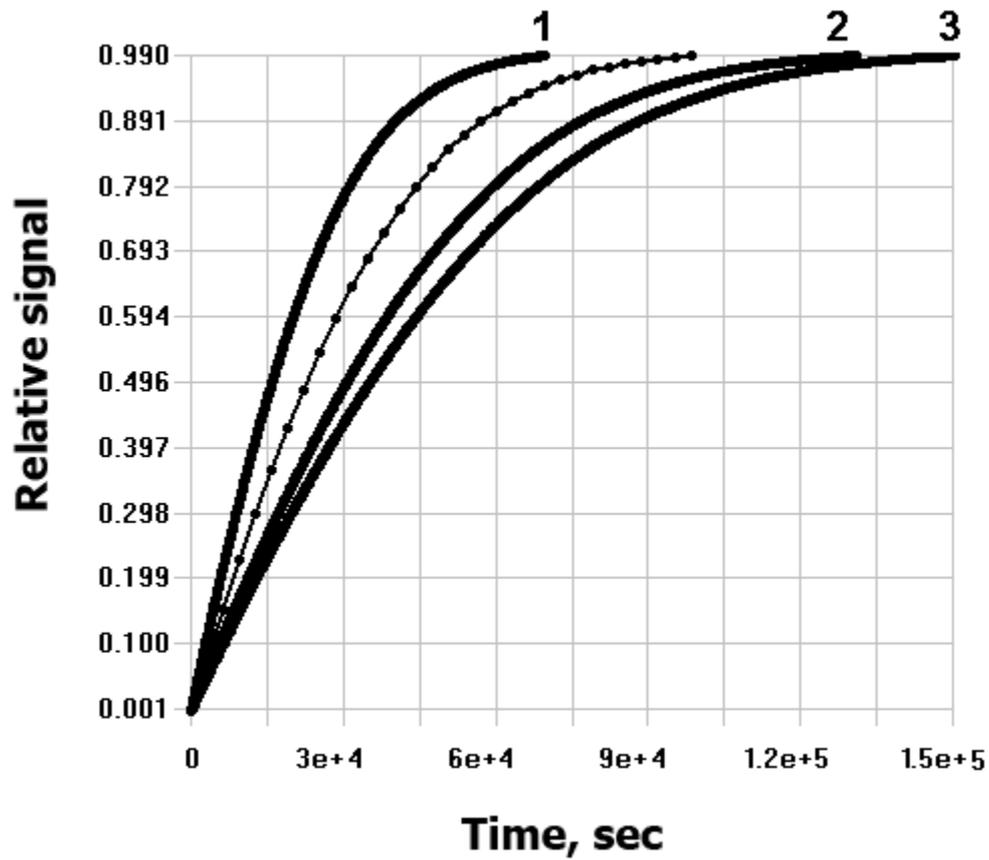

Fig. 1. The comparative kinetic curves for the saturation of relative signals, $J(t)/J_{eq}$, at the binding on hemispherical (curve with filled circles) and flat circular (all other curves) cells. The parameters for the binding kinetics on hemispherical cell were chosen to be equal to: $\eta_{eq}$ = 0.6; $\tau_B$ = 10 h = $3.6 \times 10^4$ s; and $\tau_B k_{diss}$ = 10 (see for nomenclature Eqs. (2), (11) and (12)). The parameters for the binding kinetics on flat circular cell were recalculated according to Eq. (17) for $r = 0.9R$ (curve 1); $0.5R$ (curve 2); and 0 (curve 3).